\documentclass[%
 reprint,
 amsmath,amssymb,
 aps,
 prd
]{revtex4-1}

\usepackage{graphicx}
\usepackage{dcolumn}
\usepackage{bm}

\usepackage{color}

\begin{document}


\title{The propagation delay in the timing of a pulsar orbiting a supermassive black hole}


\author{Eva Hackmann}
\email{eva.hackmann@zarm.uni-bremen.de}
\affiliation{University of Bremen, Center of Applied Space Technology and Microgravity (ZARM), Bremen, Germany}%
\affiliation{Faculty of Physics, University of Bielefeld, Postfach 100131, 33501 Bielefeld, Germany}

\author{Arnab Dhani}
\email{aud371@psu.edu}
\affiliation{%
	Pennsylvania State University, Department of Physics, Pennsylvania, USA
}%
\affiliation{
	Indian Institute of Technology Roorkee, Department of Physics, India
}
\affiliation{University of Bremen, Center of Applied Space Technology and Microgravity (ZARM), Bremen, Germany}%



\date{\today}

\begin{abstract}
The observation of a pulsar closely orbiting the galactic center supermassive black hole would open the window for an accurate determination of the black hole parameters and for new tests of General Relativity. An important relativistic effect which has to be taken into account in the timing model is the propagation delay of the pulses in the gravitational field of the black hole. Due to the extreme mass ratio of the pulsar and the supermassive back hole we use the test particle limit to derive an exact analytical formula for the propagation delay in a Schwarzschild spacetime. We then compare this result to the propagation delays derived in the usually employed post-Newtonian approximation, in particular to the Shapiro delay up to the second post-Newtonian order. For edge-on orbits we also consider modifications of the Shapiro delay which take the lensing effects into account. Our results are then used to assess the accuracy of the different orders of the post-Newtonian approximation of the propagation delay. This comparison indicates that for (nearly) edge-on orbits the new exact delay formula should be used.
		

\end{abstract}

\maketitle


\section{Introduction} \label{sec:intro}

There is strong evidence that the center of our galaxy hosts a supermassive black hole with a mass of about $4 \times 10^{6}$ solar masses \cite{Ghez2008,Gillessen2009}, known as Sagittarius A* (SgrA*). The gravitational field of this supermassive black hole can be explored by observing the motion of stars in its vicinity, which is also the method used to determine the mass of Sgr A*. This supermassive black hole is the primary target of strong efforts to test important features of black holes like the existence and shape of a back hole shadow \cite{Lu2014,Falcke2000}, the cosmic censorship conjecture \cite{Liu2012}, or the no-hair theorem \cite{Psaltis2016,Broderick2014,Liu2012}. It is also of very high relevance to constrain modified or alternative theories of gravity as well as black hole mimickers, see  \cite{Goddi2017,Hees2017,Grould2017,Cunha2015,Cunha2017,Vincent2016,Vincent2016b,Mizuno2018} and references therein. 

An exciting possibility to explore the gravitational field of Sgr A* is the existence of detectable pulsars orbiting the supermassive black hole. In the literature the number of normal neutron stars around Sgr A* with orbital periods below a hundred years was estimated to be around $1000$, and below ten years to around $100$ \cite{Pfahl2004}. These results are based on estimates related to the formation process of neutron stars. Other estimates based on non-detection are less optimistic, with about $90$ normal pulsars in the central parsec \cite{Macquart2010}. More recent analyses indicate that the population of millisecond pulsars may be very large in the galactic center \cite{Macquart2015}, with estimates of about $10000$ millisecond pulsars in the central parsec beaming towards Earth \cite{Rajwade2017}. The detection of pulsars in close vicinity of Sgr A* is a major science goal of the Square Kilometer Array \cite{Keane2014}, and of high frequency surveys of the next generation Very Large Array (ngVLA) \cite{Bower2018}. Complementary efforts to explore the spacetime close to Sgr A* are near-infrared observations on scales of a few hundred Schwarzschild radii \cite{Eisenhauer2011,Gravity2017} and very-long baseline interferometry by (sub-)mm telescopes on horizon scales \cite{Falcke2000}.

Pulsar timing gives rise to a superb determination of the physical and orbital parameters of the neutron star \cite{Verbiest2008,Kramer2012}. If a pulsar closely orbiting Sgr A* is detected, it will probably be possible to improve the accuracy of the mass estimate, and to determine the spin orientation and magnitude of the supermassive black hole \cite{Liu2012,Zhang2017}. Furthermore, tests of the no-hair theorem and the cosmic censorship conjecture will be feasible \cite{Liu2012}. Whereas in the usual binary systems the pulsar and companion mass are more or less comparable, in the considered system the mass ratio is extreme and a pulsar may orbit very close to the supermassive black holes for several years of observation times. This will induce strong relativistic effects both on the orbit and the electromagnetic radiation, which will be significantly larger than in the common binary systems, see also \cite{Wang2008,Wang2009}. In the data analysis, the relativistic effects are usually accounted for by using a set of post-Keplerian parameters, see e.g. \cite{Edwards2006,Damour1992}. This treatment is based on the work by Damour and Deruelle \cite{Damour1986}, using a post-Newtonian expansion to tackle the relativistic two body problem. As the post-Newtonian approximation assumes a weak field, the question arises if this very successful approach is reliable for the case of a pulsar which very closely orbits a supermassive black hole.

In this paper we will investigate this question for the relativistic effects on the radio pulses due to the gravitational field of the supermassive back hole. Apart from the Roemer delay, which describes the time of flight across the orbit due to the finite velocity of light, the dominant relativistic contribution is the Shapiro delay \cite{Shapiro1964,Blandford1976,Brumberg1991,Zschocke2009}, which accounts for the modified velocity of light in the curved spacetime. Other relativistic effects on electromagnetic radiation include the bending of the path in the curved background or the influence of rotation of the central supermassive black hole \cite{Schneider1990,Doroshenko1995,Wex1999}. For relativistic effects on the pulsar orbit, which we will not consider in this paper, we refer to \cite{Damour1988,Wex1995,Wex1999,Koenigsdoerfer2005,FutamaseLRR,Hartung2013,Damour2014,Levi2016,Bernard2017,Zhang2017} and reference therein. In general of course all these effects can only be disentangled in the linearised approximation of the nonlinear theory of General Relativity.      

In the considered setting of a pulsar orbiting a supermassive black hole, the extreme mass ratio justifies to consider a different approximation, namely the test particle limit. In this limit we neglect the gravitational field of the pulsar and only consider the motion of the electromagnetic radiation in the gravitational field of the supermassive black hole. As the dominant nontrivial relativistic effect on the radio pulses is the Shapiro delay, we restrict, in this paper, to the spacetime of a Schwarzschild black hole and neglect contributions due to the rotation and other (speculative) features of the central black hole. Within the setting of a Schwarzschild spacetime we calculate the exact analytical solution for the propagation delay of the pulses and compare them to the corresponding post-Newtonian approximations.

The outline of the paper is as follows. In section two we review the equations of motion for lightlike geodesics in a Schwarzschild spacetime and solve for the propagation time from a given position in spacetime to an observer at infinity. Note that the employed solution method can also straightforwardly be used to derive the time delay to an observer at a finite position. We decided to use an observer at infinity, because in pulsar timing only the differences in the delay along the pulsar orbit can be detected, and for the sake of comparison to the post-Newtonian expressions. In section three we find the finite exact propagation delay in Schwarzschild spacetime with respect to a reference point. The weak field approximations of the propagation delay are reviewed in section four, up to the second post-Newtonian order. In the fifth section we compare the exact result to the post-Newtonian approximations for a number of test cases assuming for simplicity a circular pulsar orbit. We close the paper with a summary and discussion.

\section{Null geodesics in Schwarzschild spacetime}

The Schwarzschild metric in the Boyer-Lindquist type Schwarzschild coordinates is given by
\begin{align}
ds^2 & = - \left( 1-\frac{2m}{r}\right) (dx^0)^2 + \left( 1-\frac{2m}{r}\right)^{-1} dr^2\nonumber\\
& \quad + r^2 (d\theta^2 + \sin^2\theta d\varphi^2) \label{metric}
\end{align}
where $m=\frac{GM}{c^2}$ is related to the mass $M$ of the black hole. The coordinate $x^0 = ct$ is related to the coordinate time $t$, which is due to asymptotic flatness the proper time measured by an observer at infinity. Photons propagate along null geodesics which obey the geodesic equation
\begin{align}
0 & = \ddot{x}^\mu + \Gamma^\mu_{\, \nu\rho} \dot{x}^\nu \dot{x}^\rho
\end{align}
where $\Gamma$ denotes the Christoffel symbols, and the dot indicates the derivative with respect to an affine parameter $\tau$ along the curve. Due to the spherical symmetry of the Schwarzschild spacetime geodesics remain in their orbital plane, and we may choose this orbital plane as the equatorial plane $\theta=\frac{\pi}{2}$. There are two further constants of motion due to the symmetry of the spacetime, which are related to the energy $\tilde{E}$ and the orbital angular momentum $L$ of the photon,
\begin{align}
-E & = -\tilde{E}/c = g_{00} \dot{x}^0 = -\left( 1-\frac{2m}{r}\right) \frac{cdt}{d\tau}\,,\\
L & = g_{\varphi\varphi} \dot{\varphi} = r^2 \frac{d\varphi}{d\tau}\,.
\end{align}
A third constant of motion is given by the condition for null geodesics,
\begin{align}
0 & = g_{\mu\nu} \dot{x}^\mu \dot{x}^\nu\,.
\end{align}
The equations of motion for photons in Schwarzschild spacetime are then given by
\begin{align}
\left( \frac{dr}{d\tau}\right)^2 & = E^2 - \left( 1-\frac{2m}{r}\right) \frac{L^2}{r^2} \nonumber\\
& = \frac{L^2}{r^4} \left( \frac{r^4}{b^2} - r^2 + 2mr \right) =: \frac{L^2}{r^4} R(r)  \,, \label{drdtau} \\
\frac{d\varphi}{d\tau} & = \frac{L}{r^2}\,, \label{dphidtau}\\
\frac{dx^0}{d\tau} & = \frac{E}{1-\frac{2m}{r}} \label{dtdtau}\,,
\end{align}
where $b=L/E$ is the impact parameter. Before we proceed to solve for the coordinate time let us shortly discuss eq. \eqref{drdtau}. The quartic polynomial $R(r)$ has four roots,
\begin{align}
R(r) & = \frac{r^4}{b^2} -r^2+2mr\nonumber\\
& = \frac{1}{b^2} (r-r_4)(r-r_3)(r-r_2)(r-r_2) \label{rootsR}
\end{align}
with $r_1<r_2<r_3<r_4$ if all roots are real, and $r_2=0$. Then $r_1$ is always real and negative, whereas the two positive roots $r_3$, $r_4$ will merge to a double root for $b=b_{\rm crit}=\sqrt{27}$ and become a complex pair for $b<b_{\rm crit}$. Due to the square on the left hand side of eq. \eqref{drdtau} it is clear that we have for $b>b_{\rm crit}$ either a flyby orbit, which comes from infinity turns at $r_4$ and returns to infinity, or a terminating bound orbit, which is confined to the region $0\leq r \leq r_3$. For $b<b_{crit}$ we have a terminating escape orbit, which comes from infinity and falls into the singularity at $r=0$. The critical value $b_{\rm crit}$ corresponds to the unstable circular photon orbit $r=3m$, and $b=0$ corresponds to purely radial motion. 

As we are interested in the arrival times of photons as measured by an observer at infinity we derive from eqs. \eqref{drdtau} and \eqref{dtdtau}
\begin{align}
\left(\frac{dr}{c dt}\right)^2 & = \left( 1-\frac{2m}{r}\right)^2 \frac{b^2}{r^4} R(r)\,.
\end{align}
We recast this equation in an integral form,
\begin{align}
c(t_a-t_e) & = \int_{\gamma} \frac{r^2 dr}{b \left( 1-\frac{2m}{r}\right) \sqrt{R(r)}}\\
& = \int_{r_4}^\infty \frac{r^2 dr}{b \left( 1-\frac{2m}{r}\right) \sqrt{R(r)}}\nonumber \\
& \quad \pm \int_{r_4}^{r_e} \frac{r^2 dr}{b \left( 1-\frac{2m}{r}\right) \sqrt{R(r)}} \,, \label{teq}
\end{align}
where $t_e$ corresponds to the time of emission at the radius of emission $r_e$ and $t_a$ is the time of arrival at $r=\infty$. The path of integration $\gamma$ starts at $r_e$ and then either goes to the turning point $r_4$ for the case of a flyby orbit which first decreases in radius, or directly goes to $r=\infty$. This implies that for monotonically increasing $r$ we choose the minus sign in \eqref{teq} and else the plus sign. Note that for $b<b_{\rm crit}$, $r_4$ is one of the two complex conjugate roots. The complete expression is nevertheless real.

The integral in \eqref{teq} can be solved exactly in terms of elliptic integrals. Details of the derivation can be found in appendix \ref{app:time}. The result is
\begin{align}
\frac{c}{m}(t_a-t_e) & = T(\infty,b)\pm T(r_e,b)\,, \label{soltime}
\end{align}
with $T(r,b)$ given by 
\begin{widetext}
	\begin{align}
	T(r,b) & = \frac{2}{\sqrt{r_4(r_3-r_1)}} \Bigg[ \left(\frac{r_3^3}{r_3-2}+\frac{1}{2}(r_4-r_3)(r_3-r_1+4)\right) F(x,k) - \frac{1}{2} r_4(r_3-r_1) E(x,k) - 2 (r_4-r_3) \Pi\left(x,\frac{k^2}{c_1},k\right) \nonumber\\
	& \quad  - \frac{8(r_4-r_3)}{(r_4-2)(r_3-2)} \Pi(x,c_2,k)  \Bigg] + \frac{b\sqrt{R(r)}}{r-r_3} + 2\ln\left( \frac{\sqrt{r(r-r_1)}+\sqrt{(r-r_4)(r-r_3)}}{\sqrt{r(r-r_1)}-\sqrt{(r-r_4)(r-r_3)}} \right)\,, \label{eq:solTdiv}
	\end{align}
\end{widetext}
see also \eqref{solTdiv}. Here $x$ is related to $r$ by \eqref{subs}, $k$ is defined in \eqref{defk}, $c_1,c_2$ are defined in \eqref{defAc}, and $r_1,r_2,r_3,r_4$ are the zeros of $R(r)$ as before, see \eqref{rootsR}. All quantities appearing in \eqref{eq:solTdiv} are normalised such that they are dimensionless, which is the reason for the factor $c/m$ on the left hand side in \eqref{soltime}. The Jacobian elliptic integrals $F$, $E$, and $\Pi$ are defined in \eqref{defF}-\eqref{defPi}. The (normalised) impact parameter $b$ has to be determined from the emission position, see the next section. Note that of course the time to reach infinity diverges. For a numerical integration, it is therefore not clear how to isolate the diverging parts in the integral such that they will cancel with respect to a reference point. In the analytical solution \eqref{eq:solTdiv} the last two terms cause the divergence. The diverging parts are independent of the choice of the impact parameter $b$ and of the form $r+2\ln(r)$, see \eqref{DivTaylorlin} and \eqref{DivTaylorlog}, where we provided a Taylor expansion of the diverging terms. They will therefore cancel if we consider the propagation delay with respect to a reference point.

\section{The propagation delay in Schwarzschild spacetime}

Consider a pulsar orbiting a supermassive black hole. Due to the big difference in the masses of these two objects we may consider the pulsar as a test particle and the center of mass to coincide with the center of the black hole. We choose our coordinate system $(X,Y,Z)$ with origin at the center of the black hole, and such that the $Z$--axis is defined by the line of sight from the observer to the origin. The $X$--axis is given by the ascending node of the pulsar orbit with respect to the plane of sky. We measure the inclination $i$ of the pulsar orbit then with respect to the plane of sky, which is the $X$--$Y$--plane of our coordinate system.

To calculate the time delay \eqref{teq} we need the radial coordinate $r_e$ of the pulsar and the impact parameter $b$. The latter has to be determined from the pulsar position $(r_e,\varphi_e)$ in the common plane of pulsar, observer, and black hole. We use for this the differential equation for the angle $\varphi$ from \eqref{drdtau} and \eqref{dphidtau},
\begin{align}
\left(\frac{dr}{d\varphi}\right)^2 & = R(r)\,, \quad r(\varphi_e)= r_e\nonumber\\
\Leftrightarrow \, \varphi_e & = \int_\gamma \frac{dr}{\sqrt{R(r)}} = \int_{r_4}^\infty \frac{dr}{\sqrt{R(r)}} \pm \int_{r_4}^{r_e} \frac{dr}{\sqrt{R(r)}} \,,\label{eq:phie}
\end{align} 
where we assumed that $\varphi=0$ at the observer position. The problem of solving this equation for $b$ is known as the emitter observer problem. To our knowledge there is no exact analytical solution to this. From the available analytical approximations, see Semerak \cite{Semerak2015} for a review, we tested the one by Beloborodov \cite{Beloborodov2002,DeFalco2016} which turned out to be not sufficient for some of our test cases in section \ref{sec:comp}. In general, we therefore have to solve the above equation numerically for the impact parameter $b$. For the particular test cases discussed in section \ref{sec:comp} we instead choose a grid of impact parameters and calculate the corresponding emission angles $\varphi_e$ (for fixed $r_e$) from equation \eqref{eq:phie} analytically using the Jacobian elliptic integrals introduced in the appendix. 

Let $(x,y,z)$ be the coordinate system with the pulsar in the $x$--$y$--plane, with $x=X$. As our aim is to analyse the relativistic propagation delay, we neglect the relativistic effects on the pulsar orbit and assume for simplicity a Keplerian orbit. (This assumption has no impact on our solution method, which is valid for any given emission position outside the event horizon.) Then the pulsar motion is described by $r_e=\frac{a(1-e^2)}{1+e\cos\phi}$, where $a$ is the semi major axis, $e$ is the eccentricity, and $\phi$ is the true anomaly. We find $x=r_e\cos(\omega+\phi)$, $y=r_e\sin(\omega+\phi)$, $z=0$, where $\omega$ is the argument of the periastron. Then a simple rotation along the $x$-axis by the inclination angle $i$ suffices to transform to the $(X,Y,Z)$ system. The desired angle between pulsar and observer is then given by the angle $\vartheta$ in spherical coordinates $X=r\cos\psi\sin\vartheta$, $Y=r\sin\psi\sin\vartheta$, $Z=r\cos\vartheta$. Therefore, the angle $\varphi_e$ in equation \eqref{eq:phie}, in the instantaneous common plane of pulsar and observer, is determined by $\varphi_e=\vartheta$ with $\cos\vartheta = -\sin i \sin (\omega+\phi)$ and, therefore,
\begin{align}
\cos\varphi_e = -\sin i \sin (\omega+\phi)\,.\label{relphivarphi}
\end{align}

The propagation delay is given as the difference between the time delay of a signal from the actual position of the pulsar and some reference point,
\begin{align}
\Delta_{\rm ex}(r_e,\varphi_e) & = (t_a - t_e)(r_e,\varphi_e) - (t_a - t_e)(r_{\rm ref},\varphi_{\rm ref}) \nonumber\\
& = \frac{m}{c} [T(\infty,b_e) \pm T(r_e,b_e)]\nonumber\\
& \quad - \frac{m}{c} [T(\infty,b_{\rm ref}) \pm T(r_{\rm ref},b_{\rm ref})] \label{exactdelay}
\end{align}
where $b_e=b(r_e,\varphi_e)$ and $b_{\rm ref}=b(r_{\rm ref},\varphi_{\rm ref})$. Note that the diverging terms in $T(\infty,b_e)$ and $T(\infty,b_{\rm ref})$ directly cancel each other, see \eqref{solTdiv}. In this way we only consider the finite difference between the arrival times of signals from the pulsar as it orbits the central supermassive black hole.

\section{The propagation delay in the weak field}

In the weak field approximation the propagation delay can be decomposed into several effects (where weak field means $v/c \sim m/r \ll 1$ with the velocity $v$ and the radius $r$ of the test particle). This includes the Roemer delay, which corresponds to the time of flight across the orbit, and the Shapiro delay (to the first post--Newtonian order) due to the varying three-velocity of light in the gravitational field. Usually, in the timing formula only these two effects are taken into account. Further effects are the geometric delay, which accounts for the curved path of the signal due to the spacetime curvature, and the Shapiro delay in the second post--Newtonian order.

All these effects are encoded in the fully general relativistic solution \eqref{exactdelay} derived in the foregoing section. To assess which effects should be taken into account to appropriately model a pulsar orbiting a supermassive black hole we will compare the different effects mentioned above to the general formula \eqref{exactdelay}. For the convenience of the reader we here review the different effects in the weak field approximation on the propagation delay.

\subsection{The Roemer and the Shapiro delay}

The weak field propagation delay in pulsar timing was derived by Blandford and Teukolsky \cite{Blandford1976}. We here review their results for the convenience of the reader. The Schwarzschild metric \eqref{metric} in harmonic coordinates can be approximated in the weak field regime up to the second post-Newtonian  order as \cite{Zschocke2009}
	
\begin{align}
ds^2 & = -\left(1-\frac{2m}{r} + \frac{2m^2}{r^2}\right) c^2dt^2  + \frac{m^2 x^i x^j}{r^2} dx^i dx^j \nonumber\\
& \,  + \delta_{ij} \left( 1 + \frac{2m}{r} + \frac{m^2}{r^2} \right) dx^i dx^j  \label{2PNmetric}
\end{align}
where we neglected terms of order $v^6/c^6$ and used $r=|\vec{x}|$. This can be further approximated to the first post-Newtonian order as
\begin{align}
ds^2 & = -\left(1+\frac{2\Phi}{c^2}\right) c^2dt^2 + \left(1-\frac{2\Phi}{c^2}\right) (d\vec{x})^2\,, \label{1PNmetric}
\end{align}
where we now neglected terms of order $v^3/c^3$ and used $(d\vec{x})^2 = dx^2+dy^2+dz^2$. Here $\Phi(\vec{x})=-\frac{GM}{r}$ denotes the Newtonian gravitational potential of the supermassive black hole. We will in this subsection work to the first post-Newtonian order.

The normalization condition for null geodesics, $g_{\mu\nu} dx^\mu dx^\nu = 0$, can then be written as
\begin{align}
c dt & = \sqrt{\frac{1-2\Phi/c^2}{1+2\Phi/c^2}} d\vec{x} \approx \left(1-\frac{2\Phi}{c^2}\right) d\vec{x}\,.
\end{align}
This can be integrated to
\begin{align}
c(t_a-t_e) & = \int_{\vec{r}_e}^{\vec{r}_a} \left(1-\frac{2\Phi(\vec{x})}{c^2}\right) d\vec{x}\,,
\end{align}
where $\vec{r}_e$ is the point of emission at the time of emission $t_e$, and $\vec{r}_a$ is the point of arrival at the time of arrival $t_a$. To first order it can be assumed that the signal travels on a straight line from the pulsar to the observer,
\begin{align}
\vec{x}(t) & = \vec{r}_e + \frac{t-t_e}{t_a-t_e} (\vec{r}_a-\vec{r}_e)\,.
\end{align}
Then the integral simplifies to 
\begin{align}
c(t_a-t_e) & = |\vec{r}_a-\vec{r}_e| +\frac{2GM}{c} \int_{t_e}^{t_a} \frac{dt}{|\vec{x}(t)|}\nonumber\\
& = |\vec{r}_a-\vec{r}_e| +\frac{2GM(t_a-t_e)}{c|\vec{r}_a-\vec{r}_e|}  \times \nonumber\\
& \quad \times \ln \Bigg[ \frac{\vec{r}_e(\vec{r}_a-\vec{r}_e) + |\vec{r}_a-\vec{r}_e|^2 + |\vec{r}_a||\vec{r}_a-\vec{r}_e|}{\vec{r}_e(\vec{r}_a-\vec{r}_e) + |\vec{r}_e||\vec{r}_a-\vec{r}_e|} \Bigg]\,.
\end{align}
If we now assume that $c(t_a-t_e)\approx|\vec{r}_a-\vec{r}_e|$ and $|\vec{r}_a|\gg|\vec{r}_e|$ we can further approximate the above expression,
\begin{align}
c(t_a-t_e) & =  |\vec{r}_a-\vec{r}_e| + \frac{2GM}{c^2} \ln \Bigg[ \frac{2|\vec{r}_a|}{|\vec{r}_e| + \vec{r}_e\cdot \vec{n}} \Bigg]\,, \label{PNtimesol}
\end{align}
where $\vec{n} = \vec{r}_a/|\vec{r}_a|$ is the unit vector pointing towards the observer. For the purpose of pulsar timing, we can neglect all constant contributions to the time delay, because they will only contribute by a constant shift which can not be detected. The first term in the expression \eqref{PNtimesol} corresponds to the Roemer delay. Its time varying part corresponds to the time of flight across the orbit and can be approximated as $-\vec{r}_e \cdot \vec{n}$. With $\vec{r}_e\cdot\vec{n} = -r\sin i \sin(\omega+\phi)$ we find \cite{Blandford1976}
\begin{align}
\Delta_{\rm R} & := \frac{a (1-e^2)\sin i\sin(\omega+\phi)}{c(1+e\cos\phi)}\,, \label{Roemer}
\end{align}
where we used $r:=|\vec{r}_e|=\frac{a(1-e^2)}{1+e\cos\phi}$. Here $i$ is the inclination of the orbital plane with respect to the plane of sky and $\omega$ is the argument of periapsis. For the time varying part of the second term in eq. \eqref{PNtimesol} we then find \cite{Blandford1976}
\begin{align}
\Delta_{\rm S} := \frac{2GM}{c^3} \ln \Bigg[ \frac{1+e\cos\phi}{1-\sin i \sin(\omega+\phi)} \Bigg]\,.\label{PNdelay}
\end{align}

Note that the Roemer delay vanishes at $\phi=-\omega$, whereas the Shapiro delay vanishes at
\begin{align}
\varphi = \arctan \frac{-e-\sin i \sin \omega}{\sin i \cos \omega}\,.
\end{align}
For the special case of circular orbits ($e=0$) we also find $\phi=-\omega$, i.e.~the ascending node. The point where the time delay vanishes can be considered as the reference point.

The result \eqref{PNdelay} diverges for edge--on orbits with $i=\pi/2$ at superior conjunction, $\omega+\phi=\pi/2$. This is because the assumed straight path of light passes through the central object, where we have an infinitely deep gravitational potential. To circumvent this, one can take the lensing of the path into account, which was achieved by Lai and Rafikov \cite{Lai2005}, correcting a result by Schneider \cite{Schneider1990}. The generalized result is
\begin{align}
\Delta_{\rm S,l} := \frac{2GM}{c^3} \ln \Bigg[ \frac{a(1-e^2)}{\sqrt{|\vec{r}_e\cdot \vec{n}|^2+|\vec{r}_\pm|^2}-\vec{r}_e\cdot \vec{n}} \Bigg] \label{lensedShapiro}
\end{align}
where $\vec{r}_\pm$ is the (approximate) position of the image of the source in the plane of sky,
\begin{align}
\vec{r}_{\pm} & = \frac{\vec{r}_s}{2}\left(1\pm \sqrt{1+\frac{4R_E^2}{|\vec{r}_s|^2}}\right)\,,
\end{align}
and $\vec{r}_s$ is the projection of $\vec{r}_e$ onto the plane of sky,
\begin{align}
\vec{r}_s = \vec{r}_e\sqrt{1-\sin^2i\sin^2(\omega+\phi)}\,.
\end{align}
Here $R_E$ denotes the Einstein radius, which can be approximated by $R_E^2 = \frac{4GM}{c^2} |\vec{r}_e|\sin i$ at superior conjunction $|\vec{r}_e|=a(1-e^2)/(1+e\sin\omega)$.

\subsection{The geometric delay}

The geometric delay is the extra time that the light ray takes due to the curved path it takes in a gravitational potential. This delay is taken into account only when the pulsar is on the farther side of the orbit relative to the black hole. To first order the path taken by the light ray is considered to be a straight line from its point of emission to its minimum distance to the black hole and from there to the observer. The delay is the difference between this path length to the straight line path from the pulsar to the observer. 

This delay to first order is given by \cite{Lai2005}
\begin{align}
\Delta_{\rm geo} = \frac{2GM}{c^3} \Bigg[ \frac{|\vec{r}_{\pm}-\vec{r}_s|}{R_E} \Bigg]^2\,. \label{geodelay}
\end{align}
As pointed out by Lai and Rafikov \cite{Lai2005}, if $R_E$ is large compared to $|\vec{r}_s|$ we have $|\vec{r}_{\pm}-\vec{r}_s|\to R_E$ and the two images merge into an Einstein ring. In the opposite case, $|\vec{r}_s| \gg R_E$ we find $|\vec{r}_{+}-\vec{r}_s|\to R_E^2/|\vec{r}_s|$ and $|\vec{r}_{-}-\vec{r}_s|\to |\vec{r}_s|$, but the "--" image is very faint and its contribution is negligible.

The geometric delay is most significant when the pulsar is directly behind the black hole.


\subsection{The second order Shapiro delay}

The second post--Newtonian order of the Shapiro delay can be derived, for instance, from a result of Zschocke and Klioner \cite{Zschocke2009}. From their eq. (24) we find the formula
\begin{align}
\Delta_{\rm 2PN}= \frac{GM}{c^3r} \left[ -\frac{4}{1+\cos\varphi_e}  + \frac{\cos\varphi_e}{4} + \frac{15\varphi_e}{4\sin\varphi_e} \right]
\end{align}
where $\varphi_e$ is the angle between the emission position vector $\vec{r}_e$ and receiver position vector $\vec{r}_a$, as before. This implies $\cos\varphi_e = -\sin i \sin(\omega+\phi)$ in our notation. Here $r = a(1-e^2)/(1+e\cos\phi)$ as before.

Note that at inferior conjunction $\varphi=0$ the last term in the above equation does not diverge, whereas at superior conjunction $\varphi=\pi$ the first term shows the familiar divergence. However, this time the delay goes to minus infinity.

Note that the delays derived in this section are in general coordinate dependent quantities, which we have to keep in mind for a comparison.  



\section{Comparison of exact and post-Newtonian propagation delay} \label{sec:comp}

In this section we will compare the two different approaches to calculate the relativistic propagation delay derived or reviewed in this paper, and use this comparison to access the quality of the post-Newtonian approximations for the case under discussion here: a pulsar, considered as test particle, orbiting a supermassive black hole. For the mass of the black hole we assume in the following $4\times 10^6$ solar masses with $m_{\rm Sun}=\frac{GM_{\rm Sun}}{c^2} = 1476$ meter. We will present our result for the propagation delay in seconds. The dimensionless value can be recovered by dividing by $GM/c^3 \approx 19.7\, \rm sec$. For another black hole mass $M_2$ our results will be rescaled by a factor $M_2/M$.

The post-Newtonian metric \eqref{1PNmetric} is given in harmonic coordinates, which are related to the Boyer-Lindquist type coordinates of the Schwarzschild metric \eqref{metric} by $r_{\rm PN} = r-m$. However, in General Relativity coordinates do not have any intrinsic physical meaning, what makes it difficult to compare results derived in different spacetimes. Consider a circular orbit with Schwarzschild radial coordinate $r$. An invariant characteristic of this orbit is its circumference as measured with the metric \eqref{metric}, and we could use this as the defining feature of the orbit. For the Schwarzschild metric we find the circumference of a circular orbit as $2\pi r$. In the first order post-Newtonian metric \eqref{1PNmetric} we find for the circumference of a circular orbit of radius $r_{\rm PN}$ the expression $2\pi r_{\rm PN} \sqrt{1+\frac{2m}{r_{\rm PN}}}$. Fixing the circumference of the orbit therefore gives the relation $r_{\rm PN} = \sqrt{m^2+r^2}-m$. In the limit of large radii this coincides with $r_{\rm PN} = r-m$, but for finite radii there is a small difference.

\begin{figure}
	\includegraphics[width=0.4\textwidth]{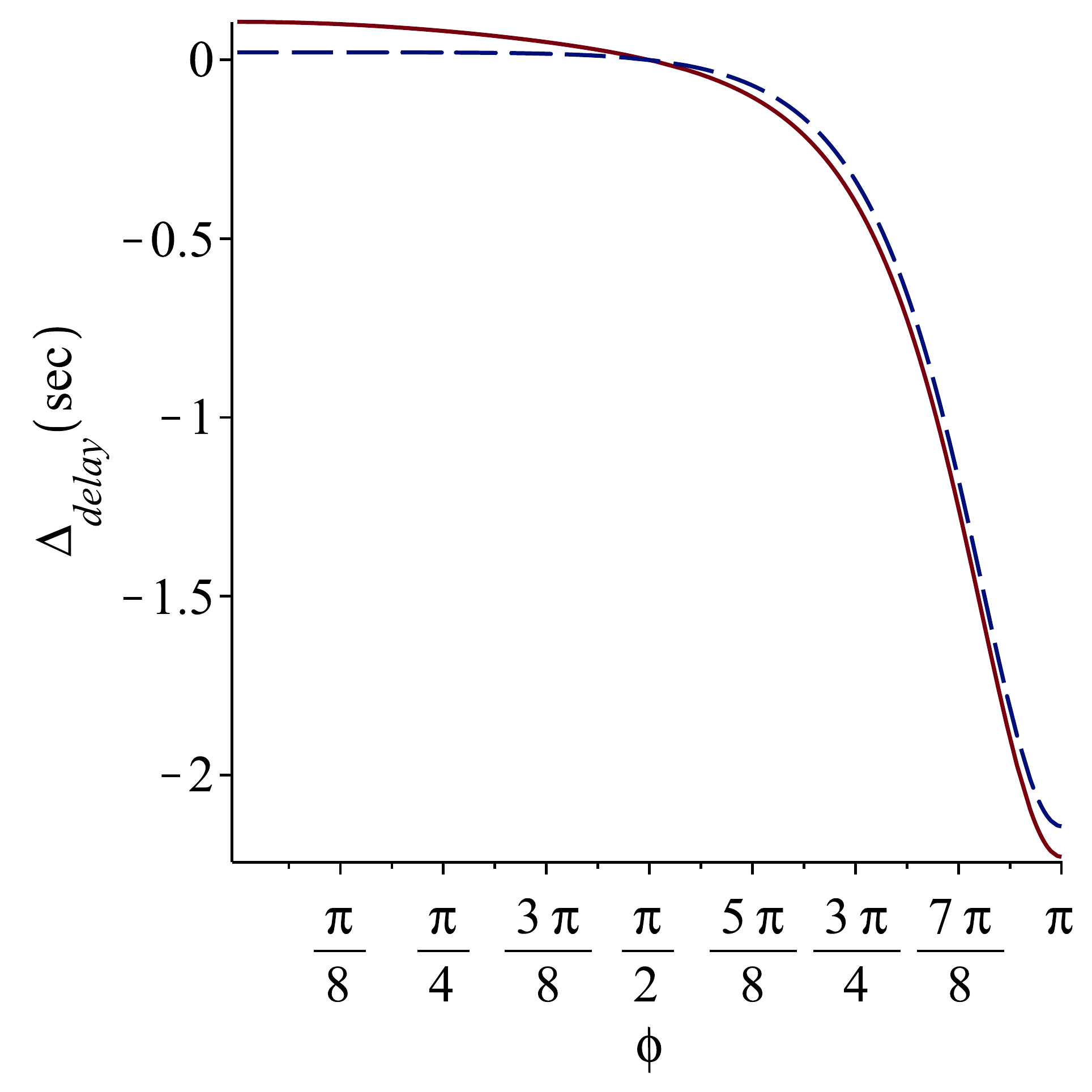}
	\caption{Difference $\Delta_{\rm delay} = \Delta_{\rm ex}-\Delta_{\rm R}-\Delta_{\rm S}$ of the Shapiro delay in Schwarzschild spacetime and the post-Newtonian approximation for a circular orbit with Schwarzschild radial coordinate $r=100m$ and inclination $\theta=\pi/3$. For the solid red line we used $r_{\rm PN}=\sqrt{m^2+r^2}-m$, for the dashed blue line $r_{\rm PN}=r-m$.}
	\label{Fig:CircHarmonicDiff}
\end{figure}

\begin{figure}
	\centering
	\includegraphics[width=0.4\textwidth]{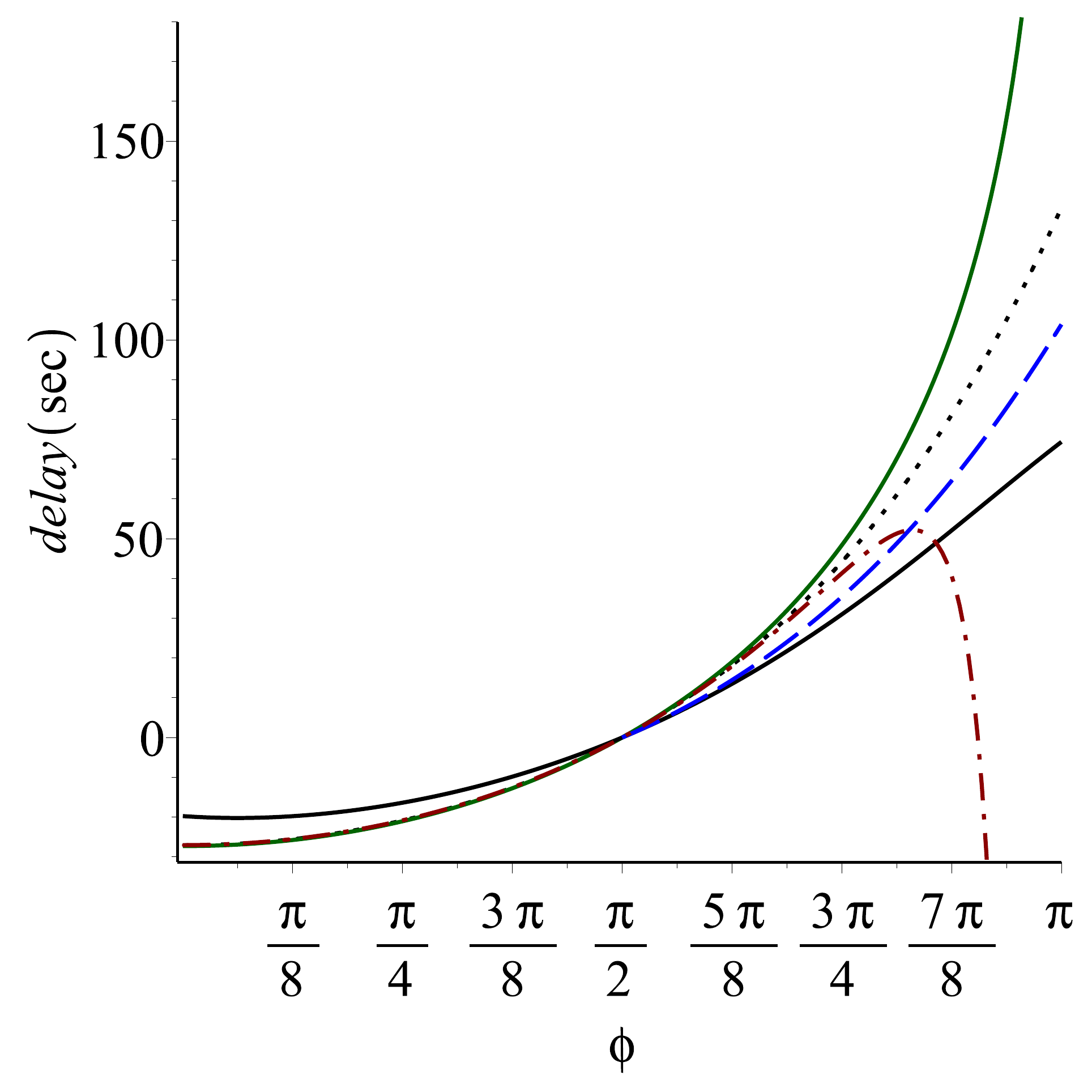}
	\caption{Comparison of post-Newtonian formulas of the Shapiro delay and the exact Schwarzschild result for a circular edge-on orbit with Schwarzschild radial coordinate $r=10m$. The black dotted line is $\Delta_{\rm ex}-\Delta_{\rm R}$, the green solid line denotes $\Delta_{\rm S}$, the black solid line $\Delta_{\rm S,l}$, the blue dashed line $\Delta_{\rm S,l}+\Delta_{\rm geo}$, and the red dash dotted line $\Delta_{\rm 2PN}$. Note that the green line diverges at $\phi=\pi$.}
	\label{Fig:edgeonshape}
\end{figure}


In figure \ref{Fig:CircHarmonicDiff} we show a plot of the difference between the exact propagation delay $\Delta_{\rm ex}$ in Schwarzschild spacetime and the Roemer delay $\Delta_{\rm R}$ plus the first order post-Newtonian approximation of the Shapiro delay $\Delta_{\rm S}$. We consider here a circular orbit with a Schwarzschild radial coordinate of $r=100m$ and an inclination of $\pi/3$ with respect to the plane of sky. As reference point we choose the ascending node with respect to the plane of sky, i.e. $\varphi_{\rm ref}=\pi/2$. With the choice $\omega=-\pi/2$ this results in $\phi_{\rm ref}=\pi/2$, see also \eqref{relphivarphi}. To achieve that the different time delays vanish at this point we add global constant offsets if necessary. Note that the first order post-Newtonian Shapiro delay $\Delta_{\rm S}$ is independent of the choice of the radius, but the Roemer delay $\Delta_{\rm R}$ depends linearly on the radius. In the plot we compare the two identifications $r_{\rm PN} = r-m$ and $r_{\rm PN} =\sqrt{m^2+r^2}-m$. From the plot it is obvious that the post-Newtonian expressions give slightly better results if we choose to identify the harmonic coordinates in the two spacetimes. Therefore, we use from now on the identification $r_{\rm PN}=r-m$ if not explicitly stated otherwise.

Note that there are further possibilities to compare the results in the Schwarzschild metric \eqref{metric} and the post-Newtonian metric \eqref{1PNmetric}, for instance we could choose the proper orbital period along the circular orbit as defining feature. However, the identification $r_{\rm PN}=r-m$ works best for all possibilities we tested.  

\begin{figure*}
	\centering
	\includegraphics[width=0.4\textwidth]{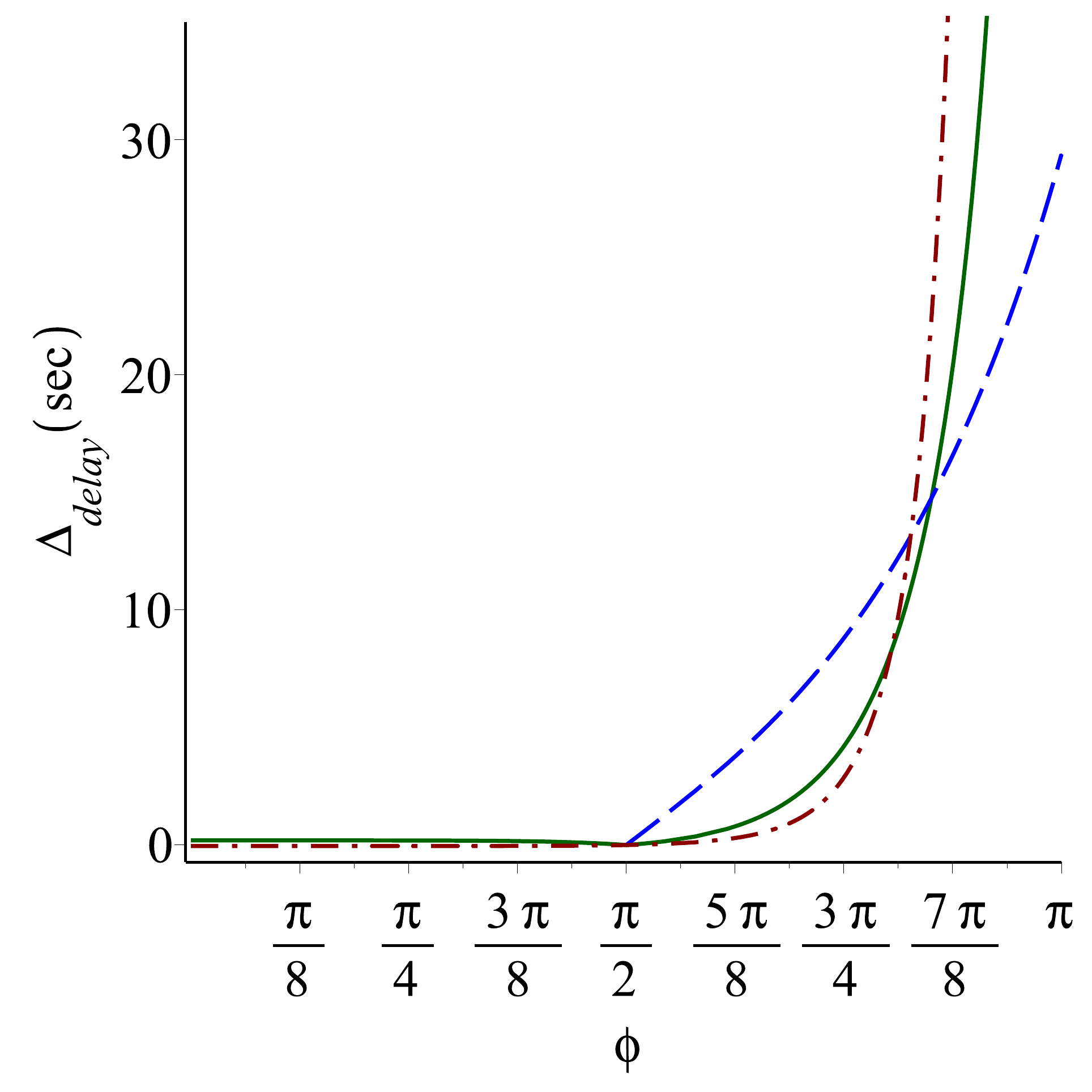}\quad
	\includegraphics[width=0.4\textwidth]{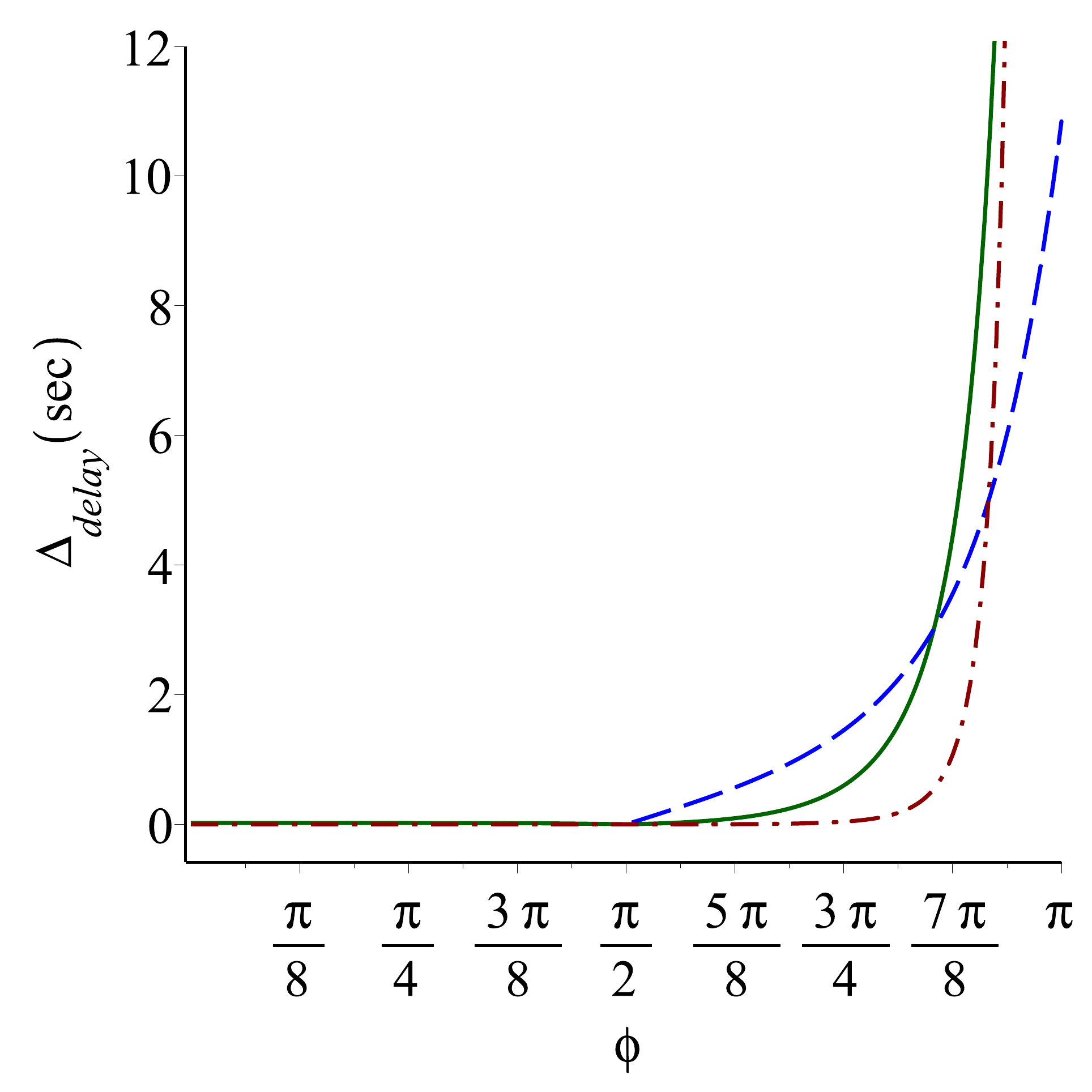}
	\caption{Absolute difference of the exact delay and the usual post-Newton delay $|\Delta_{\rm ex}-\Delta_{\rm R}-\Delta_{\rm S}|$ (solid green line) compared to the lensed delay including the geometric delay $|\Delta_{\rm ex}-\Delta_{\rm R}-\Delta_{\rm S,l}-\Delta_{\rm geo}|$ (dashed blue line) and the second order delay $|\Delta_{\rm ex}-\Delta_{\rm R}-\Delta_{\rm S}-\Delta_{\rm 2PN}|$ (dash dotted red line) for circular edge-on orbits with different Schwarzschild radii. Left: $r=10m$. Right: $r=100m$.}
	\label{Fig:DeltaDiff}
\end{figure*}

The relativistic effects are expected to be strongest if we encounter somewhere along the pulsar orbit a superior conjunction, where the pulsar is directly behind the black hole on the line of sight to the observer. In this case the usual first order post-Newtonian Shapiro delay $\Delta_{\rm S}$ diverges, as the light bending effect is neglected and the time delay is calculated as if the signal would pass through the singularity. Therefore it seems appropriate to use a modified formula taking into account the lensing and the geometric delay, as the one derived by Lai and Rafikov \cite{Lai2005}, see also \eqref{lensedShapiro} and \eqref{geodelay}. To asses the quality of the different post-Newtonian formulas for this case of an edge-on orbit, we will compare their predicted time delay to the exact result in Schwarzschild spacetime. For simplicity we use a set of circular orbits with increasing Schwarzschild radial coordinates. As the reference point we use again the ascending node with respect to the plane of sky, i.e.~$\varphi_{\rm ref}=\pi/2$, which with $\omega=-\pi/2$ results in $\phi_{\rm ref} = \varphi_{\rm ref}=\pi/2$, see \eqref{eq:phie}. At this point the Roemer delay $\Delta_{\rm R}$ as well as the usual first order Shapiro delay $\Delta_{\rm S}$ vanish, but the modified delay $\Delta_{\rm S,l}+\Delta_{\rm geo}$ as well as the second order delay $\Delta_{\rm 2PN}$ show small offsets. The exact result $\Delta_{\rm ex}$ also shows a (considerable) offset. We corrected all these offsets by adding global constants to the individual delays such that they exactly vanish at $\phi_{\rm ref}$. 

\begin{table}
	\begin{tabular}{c|c|c|c|c}
		& $r=10m$ & $r=10^2m$ & $r=10^3m$ & $r=10^4m$ \\
		& $P\approx 0.93\rm h$ & $P\approx 1.41 \rm d$ & $P\approx 0.12 \rm y$ & $P\approx 3.92 \rm y$\\
		\hline
		delay (sec) & $29.4$ & $10.8$ & $3.6$ & $1.2$  \\
		delay (dim.-less) & $1.49$ & $0.55$ & $0.18$ & $0.06$ \\
		rel.~error & $0.10$ & $5.1 \times 10^{-3}$ & $1.8 \times 10^{-4}$ & $5.8 \times 10^{-6}$  
	\end{tabular}
	\caption{Maximal difference $\Delta_{\rm ex}-\Delta_{\rm R}-\Delta_{\rm S,l}-\Delta_{\rm geo}$ for circular edge on orbits with different radii $r$ and corresponding Keplerian orbital period $P$. The first line gives the delay in seconds for the assumed $4\times 10^6$ solar mass black hole. The second line is this delay divided by $GM/c^3 \approx 19.7\, \rm sec$, which is dimensionless. The third line gives the maximal relative error of the post-Newtonian approximation $\Delta_{\rm R}+\Delta_{\rm S,l}+\Delta_{\rm geo}$.}
	\label{Tab:diff}
\end{table}

\begin{figure}
	\centering
	\includegraphics[width=0.4\textwidth]{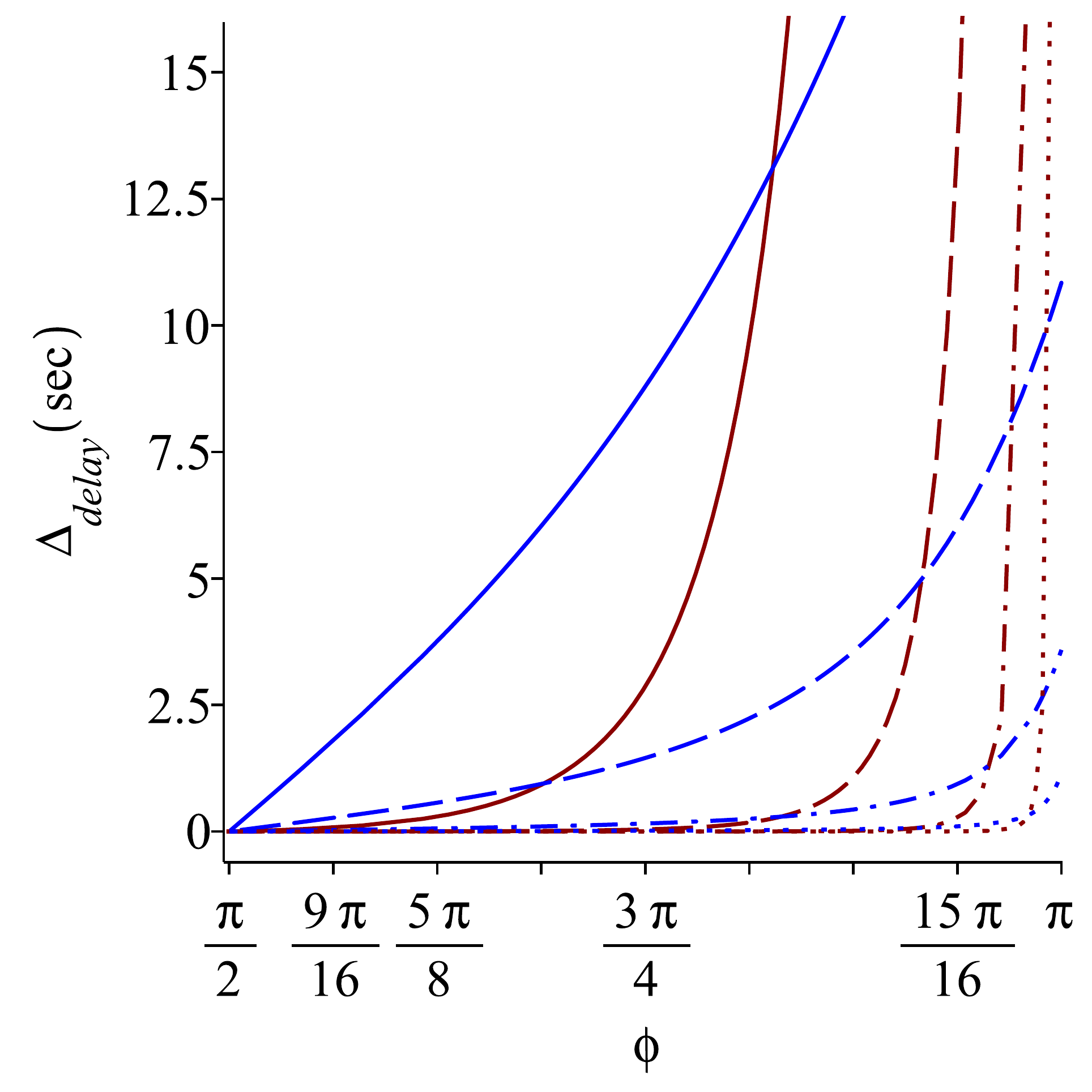}
	\caption{Comparison of the absolute differences $|\Delta_{\rm ex}-\Delta_{\rm R}-\Delta_{\rm S}-\Delta_{\rm 2PN}|$ (red lines) and $|\Delta_{\rm ex}-\Delta_{\rm R}-\Delta_{\rm S,l}-\Delta_{\rm geo}|$ (blue lines) for circular edge-on orbits with different Schwarzschild radial coordinates. For the solid lines $r=10m$, for the dashed lines $r=10^2m$, for the dash-dotted lines $r=10^3m$, and for the dotted lines $r=10^4m$.}
	\label{Fig:DeltaDiffvarR}
\end{figure}

\begin{figure*}
	\includegraphics[width=0.4\textwidth]{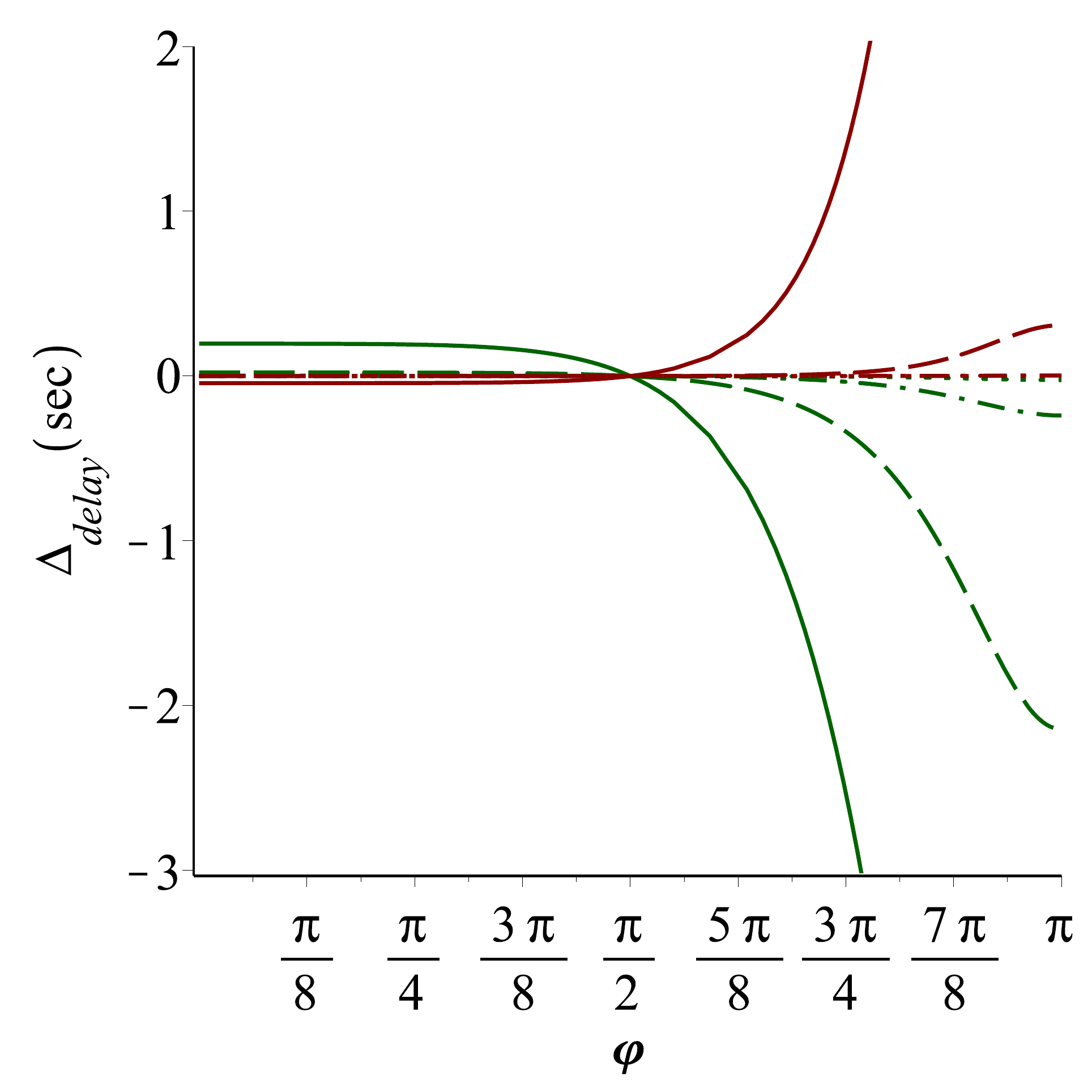}\qquad
	\includegraphics[width=0.4\textwidth]{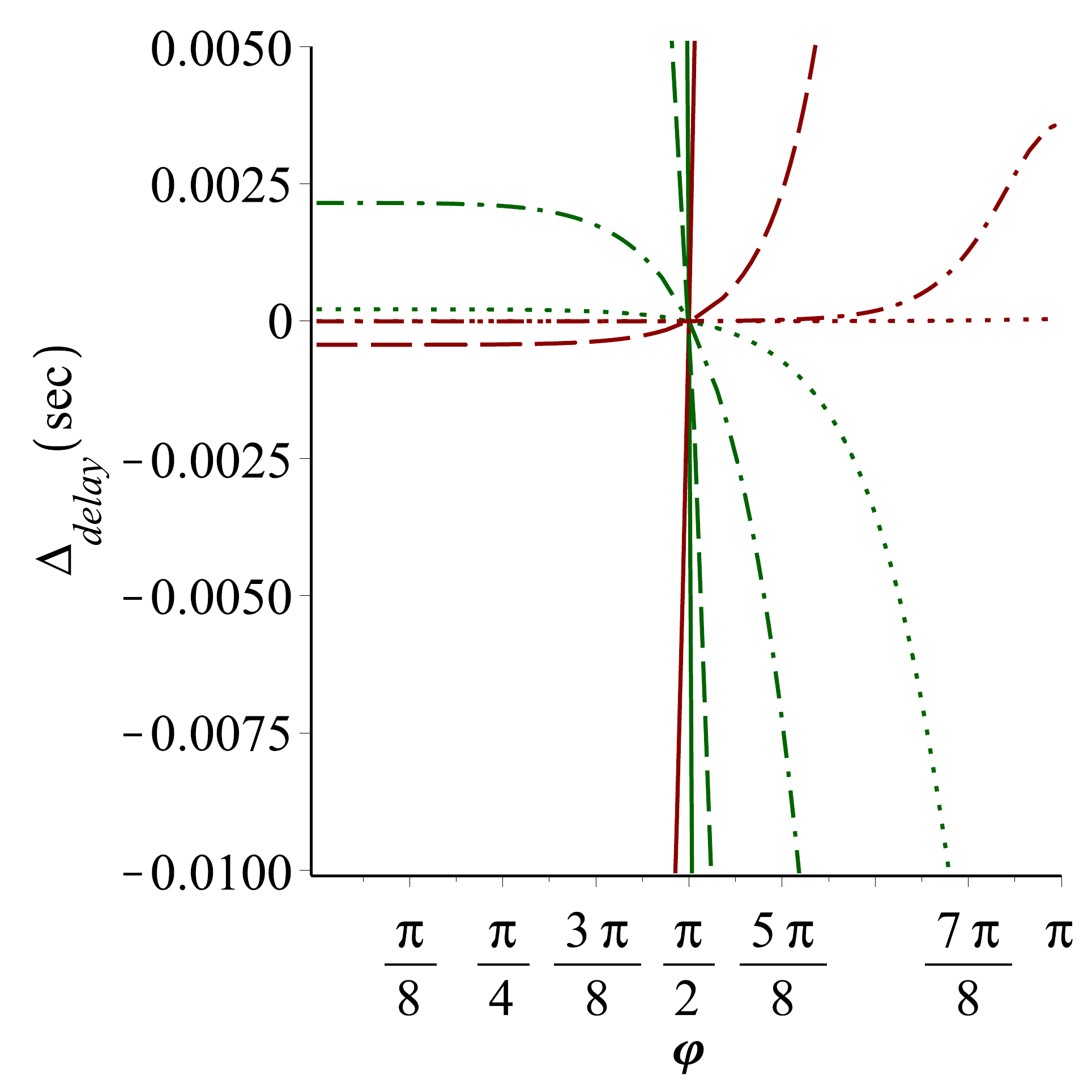}
	\caption{Difference of the first and second order post-Newtonian Shapiro delay to the exact delay for a circular orbit with inclination $\pi/3$ and various radii. For solid lines $r=10m$, for dashed lines $r=10^2m$, for dash dotted $r=10^3m$, and for dotted lines $ r=10^4m$. The green lines with negative differences at $\varphi=\pi$ correspond to $\Delta_{\rm ex}-\Delta_{\rm R}-\Delta_{\rm S}$, and the red lines with positive differences at $\varphi=\pi$ correspond to $\Delta_{\rm ex}-\Delta_{\rm R}-\Delta_{\rm S}-\Delta_{\rm 2PN}$. The right plot is a zoom of the left plot.}
	\label{Fig:inclined}
\end{figure*}

In figure \ref{Fig:edgeonshape} we plotted a typical shape of the Shapiro delay. To clearly illustrate our results we chose a unrealistically small radius of the pulsar, with an extremely short lifetime, for the sake of comparison. It can be seen that the lensed Shapiro delay plus the geometric delay behaves perfectly regular at superior conjunction $\phi=\pi$, but is still quite far away from the exact reference for the chosen radius of $r=10m$. This is also shown in figure \ref{Fig:DeltaDiff}, where we plotted the differences of the two formulas to the exact result (minus the Roemer delay). In particular, we see that the lensed plus the geometric delay is actually worse than the usual first or second order post-Newtonian formula in some region right of $\phi=\pi/2$. The maximal deviation from the exact result is reached at $\phi=\pi$. The dependence of the accuracy of the post-Newtonian Shapiro delay on the radius of the pulsar orbit is explored figure \ref{Fig:DeltaDiffvarR}. We increased the Schwarzschild radial coordinate of the circular orbit for every linestyle by a factor of ten. As expected, the difference between the exact delay and the post-Newtonian approximations reduces with increasing radius. We summarize the maximal deviation of the lensed delay including the geometric delay for the considered cases in table \ref{Tab:diff}. For a radius of $10^4 m$, which corresponds to a Keplerian orbital period of about $3.9$ years, there is still a moderate difference of about $1.2$ seconds at superior conjunction for the assumed supermassive black hole of $4\times 10^6$ solar masses. Note that for this example the light ray passes the back hole at about $200m$, which implies that the first order post-Newtonian correction should roughly be accurate to $m/r=0.005$, compared to the actual error of $0.06$. However, $m/r$ only gives an order of magnitude estimate which does not take into account any (possibly large) prefactors.

We conclude that for edge on orbits close to superior conjunction the usual Shapiro delay formulas are diverging and therefore are useless. The difference between the lensed Shapiro delay and the exact delay, $\Delta_{\rm ex}-\Delta_{\rm R}-\Delta_{\rm S,l}-\Delta_{\rm geo}$ are still quite significant at superior conjunction. For slightly inclined orbits the angle $\phi$ along the pulsar orbit shown in figure \ref{Fig:edgeonshape} will no longer coincide with the angle $\varphi$ between pulsar and observer. In this case the range of $\varphi$ will be smaller and will in particular not reach $\pi$. From figure \ref{Fig:DeltaDiffvarR} we can conclude that we should apply the lensed formula only if the pulsar is either far enough from the black hole such that the difference of the usual and the lensed delay is marginal away from superior conjunction or the inclination is so small that the usual Shapiro delay cannot be used to fit the data. For (nearly) edge-on pulsars with a high timing precision our results indicate the it is more appropriate to use the exact timing formula \eqref{exactdelay}. Of course we neglected spin effects so far, which might be bigger than the difference to the post-Newtonian treatment. We foresee however no major difficulties to generalise our results to Kerr spacetime.

Let us now turn to inclined orbits. For this case the usual first order post-Newtonian Shapiro delays $\Delta_{\rm S}$ and $\Delta_{\rm 2PN}$ are regular along the complete orbit of the particle. We want to test the accuracy of the (second order) post-Newtonian approximation for this case. For this we plotted in figure \ref{Fig:inclined} the differences between the exact delay and the post-Newtonian expressions for a circular orbit with inclination $\pi/3$. We increased the radius of the circular orbit for every linestyle by a factor of ten. For instance, we see that for a radius of $r=10^3m$, which corresponds to an Keplerian orbital period of about $0.12$ years, the first order post-Newtonian approximation still deviates by $\approx 2.3 \times 10^{-1}$ seconds, whereas the second post-Newtonian approximation is about two orders of magnitude better with a deviation of only $\approx 3.6 \times 10^{-3}$ seconds for the assumed supermassive back hole of $4\times 10^6$ solar masses. Note that for a supermassive black hole with a different mass the result only scales accordingly. For convenience, we collected the minimal and maximal differences for the cases under discussion here in table \ref{Tab:inclined}. 

\begin{table}
	\begin{tabular}{c|c|c|c|c}
		& $r=10m$ & $r=10^2m$ & $r=10^3m$ & $r=10^4m$ \\
		\hline
		min 1PN & $-11.83$ & $-2.14$ & $-2.39 \times 10^{-1}$ & $-2.42 \times 10^{-2}$  \\
		max 1PN & $1.96 \times 10^{-1}$ & $2.13\times 10^{-2}$ & $2.15 \times 10^{-3}$ & $2.15 \times 10^{-4}$ \\
		\hline
		rel.~error & $0.051$ & $1.2 \times 10^{-3}$ & $1.4 \times 10^{-5}$ & $1.4 \times 10^{-7}$ \\
		\hline
		\hline
		min 2PN & $-4.30 \times 10^{-2}$ & $-4.30 \times 10^{-4}$ & $-4.31 \times 10^{-6}$ & $-4.31 \times 10^{-8}$\\
		max 2PN & $15.15$ & $3.09 \times 10^{-1}$ & $3.58 \times 10^{-3}$ & $3.65 \times 10^{-5}$\\
		\hline
		rel.~error & $0.074$ & $1.7 \times 10^{-4}$ & $2.1 \times 10^{-7}$ & $2.1 \times 10^{-10}$
	\end{tabular}
\caption{Maximal and minimal differences between the exact delay and the post-Newtonian approximations for a circular orbit with inclination $\pi/3$ and different radii. The two lines min (max) 1PN show the minimum (maximum) of $\Delta_{\rm ex}-\Delta_{\rm R}-\Delta_{\rm S}$ in seconds. This can be converted to dimensionless values by dividing by $GM/c^3 \approx 19.7\, \rm sec$. The third line gives the maximal relative error of the 1PN approximation. The two lines min (max) 2PN show the minimum (maximum) of $\Delta_{\rm ex}-\Delta_{\rm R}-\Delta_{\rm S}-\Delta_{\rm 2PN}$ in seconds, which can again be converted into dimensionless values with the factor $GM/c^3$. The last line gives the maximal relative error of the 2PN approximation. The extremal values are always reached at $\phi=0,\pi$ for the chosen reference point at $\phi_{\rm ref}=\pi/2$.}
\label{Tab:inclined}
\end{table}

%
%
%
%

\section{Summary and outlook}

In this paper we derived an exact analytical solution for the time delay of lightlike geodesics in Schwarzschild spacetime in terms of Jacobian elliptic integrals. By isolating the diverging parts we were able to find an explicit analytical formula for the finite propagation delay with respect to a reference point. This result can be interpreted as the relativistic propagation delay of the signals of pulsars orbiting a supermassive black hole, where the extreme mass ration justifies to consider the pulsar as a test particle. We then compared our result to known post-Newtonian expressions for the propagation delay in pulsar timing: the Roemer delay, the first and second order Shapiro delay, the first order Shapiro delay including lensing effects, and the geometric delay. We discussed a suitable method to compare these results, which were derived in different spacetime, and found that an identification of the harmonic coordinates in both spacetimes yields the best results. For the comparison of the exact result and the post-Newtonian approximations, we chose a common reference point and added to each individual delay a constant offset such that it vanishes at the reference point. Note that the Shapiro delay for a pulsar closely orbiting Sgr A*, say in a $\sim$ 4 year orbit, can easily vary by about $100$ seconds along the orbit. This may affect pulsar searches in the galactic center, and further research in how the amplitude of this delay could affect observed pulse periods and therefore the sensitivity of pulsar-search algorithms to pulsars orbiting Sag A*, is warranted.

We then explored the accuracy of the post-Newtonian approximation using a number of test cases, including edge-on orbits and inclined orbits. Although we only tested circular pulsar orbits for simplicity, our method works just as well for elliptical (or more general) pulsar orbits. Our results give a clear benchmark of the accuracy of the post-Newtonian approximations and can be used as a guide to decide whether to employ a certain order of the post-Newtonian approximation. For (nearly) edge-on orbits, we showed that the lensed Shapiro delay \cite{Lai2005} still quite significantly deviates from the actual propagation delay. In this case it is more appropriate to use the presented exact delay formula \eqref{exactdelay}.

A natural continuation of our work would be to include the rotation of the central supermassive black holes in the calculations. A pulsar which closely orbits Sgr A* should in general be sensitive to the frame dragging effects \cite{Liu2012}. An additional difficulty is then that the pulses will not stay in a single plane. However, in principle this problem is also solvable in terms of elliptic integrals. Another direction to continue this work would be to also test higher order post-Newtonian approximations of the propagation delay.

\section*{Acknowledgment}
The authors thank the research training group GRK 1620 "Models of Gravity", funded by the German Research Foundation (DFG), for support. E.H. gratefully acknowledges support from the DFG funded collaborative research center SFB 1128 "Relativistic geodesy with quantum sensors (geo-Q)". A.D.~is thankful to University of Bremen for its hospitality and support where this work was conceptualized. We thank D.~Schwarz and J.~Verbiest for fruitful discussions.

\appendix

\section{Solution to the time integral}\label{app:time}

The integrals in eq. \eqref{teq} are of the form
\begin{align}
T(r,b) := \int_{r_4}^r \frac{r^2 dr}{b \left( 1-\frac{2}{r}\right) \sqrt{R(r)}}
\end{align}
with $r=r_e$ or $r=\infty$, and $R$ is defined in eq. \eqref{drdtau}. Here we normalised all quantities such that they are dimensionless, i.e.~$r=\tilde{r}/m$, $b=\tilde{b}/m$ where the twiddled quantities are in geometrised units. This integral can be analytically solved in terms if elliptic integrals. The substitution 
\begin{align}
x^2 & = \frac{(r-r_4)(r_3-r_1)}{(r-r_3)(r_4-r_1)} \label{subs}
\end{align}
with the roots $r_i$ of $R$ chosen as in eq. \eqref{rootsR}, casts the integral in the Legendre form
\begin{align}
T(r,b) & = \frac{2}{\sqrt{r_4(r_3-r_1)}} \int_{0}^{x(r)} \frac{f(x) dx}{\sqrt{(1-x^2)(1-k^2x^2)}}\,,
\end{align}
where 
\begin{align}
k^2 & = \frac{r_3(r_4-r_1)}{r_4(r_3-r_1)}\,, \label{defk}\\
f(x) & = \frac{r_3^3}{r_3-2} + \frac{A_1}{1-c_1x^2} + \frac{A_2}{1-c_2x^2} + \frac{A_3}{(1-c_3x^2)^2}
\end{align}
with the constants
\begin{align}
A_1 & = 2(r_4-r_3)(r_3+1)\,, & c_1 & = \frac{r_4-r_1}{r_3-r_1}\,,\nonumber\\
A_2 & = \frac{8(r_3-r_4)}{(r_3-2)(r_4-2)}\,, & c_2 & = \frac{(r_4-r_1)(r_3-2)}{(r_3-r_1)(r_4-2)} \label{defAc}\\
A_3 & = (r_4-r_3)^2\,, & c_3 & = c_1 \nonumber\,.
\end{align}
We note that for $r=\infty$ eq.~\eqref{subs} reduces to\\
\begin{align}
x_{\infty}^2 & := x(r=\infty)^2 = \frac{r_3-r_1}{r_4-r_1} = \frac{1}{c_1}\,.
\end{align}
With the Jacobi elliptic integrals
\begin{align}
F(x,k) & = \int_0^x \frac{dx}{\sqrt{(1-x^2)(1-k^2x^2)}}\,, \label{defF}\\
E(x,k) & = \int_0^x \sqrt{1-k^2x^2} dx\,, \label{defE}\\
\Pi(x,c,k) & = \int_0^x \frac{dx}{(1-cx^2)\sqrt{(1-x^2)(1-k^2x^2)}} \,, \label{defPi}
\end{align}
we find
\begin{widetext}
\begin{align}
T(r,b) & = \frac{2}{\sqrt{r_4(r_3-r_1)}} \Bigg[ \frac{r_3^3}{r_3-2} F(x,k) + A_1 \Pi(x,c_1,k) + A_2 \Pi(x,c_2,k) + \frac{A_3}{2(c_3-1)} \Bigg( \frac{x c_3^2\sqrt{(1-x^2)(1-k^2x^2)}}{(1-c_3x^2)(c_3-k^2)}\nonumber\\
& \quad + F(x,k)-\frac{c_3}{c_3-k^2} E(x,k) + \frac{c_3^2+3k^2-2c_3(1+k^2)}{c_3-k^2} \Pi(x,c_3,k)  \Bigg)  \Bigg]\nonumber\\
& = \frac{2}{\sqrt{r_4(r_3-r_1)}} \Bigg[ \left(\frac{r_3^3}{r_3-2}+\frac{(r_4-r_3)(r_3-r_1)}{2}\right) F(x,k) - \frac{1}{2} r_4(r_3-r_1) E(x,k) + 2 (r_4-r_3) \Pi(x,c_1,k)\nonumber\\
& \quad - \frac{8(r_4-r_3)}{(r_4-2)(r_3-2)} \Pi(x,c_2,k)  \Bigg] + \frac{b\sqrt{R(r)}}{r-r_3}\,, \label{Tsol}
\end{align}
\end{widetext}
where $x$ is related to $r$ via \eqref{subs}. Note that the Jacobi elliptic integrals can be evaluated without using a numeric integration, and can therefore be considered as an exact analytical solution to the integral $T$. 

Note that the last term in \eqref{Tsol} diverges linearly for $r  \to \infty$. As well, $\Pi(x,c,k)$ diverges logarithmically for $x^2=1/c$, which happens in our case for $x=x_\infty$ and $c=c_1=c_3$. Therefore, the time for reaching $r=\infty$ diverges as expected. To isolate the diverging terms we apply an identity,
\begin{align}
\Pi(x,c,k) & = F(x,k) - \Pi\left(x,\frac{k^2}{c},k\right) + \frac{\ln(Z)}{2P} 
\end{align}
where
\begin{align}
Z & = \frac{\sqrt{(1-x^2)(1-k^2x^2)}+Px}{\sqrt{(1-x^2)(1-k^2x^2)}-Px}\,,\\
P^2 & = \frac{(c-1)(c-k^2)}{c} = \frac{(r_4-r_3)^2}{r_4(r_3-r_1)}\,.
\end{align}
for $c=c_1$. With this equation \eqref{Tsol} becomes
\begin{widetext}
\begin{align}
T(r,b) & = \frac{2}{\sqrt{r_4(r_3-r_1)}} \Bigg[ \left(\frac{r_3^3}{r_3-2}+\frac{1}{2}(r_4-r_3)(r_3-r_1+4)\right) F(x,k) - \frac{1}{2} r_4(r_3-r_1) E(x,k) - 2 (r_4-r_3) \Pi\left(x,\frac{k^2}{c_1},k\right) \nonumber\\
& \quad  - \frac{8(r_4-r_3)}{(r_4-2)(r_3-2)} \Pi(x,c_2,k)  \Bigg] + \frac{b\sqrt{R(r)}}{r-r_3} + 2\ln\left( \frac{\sqrt{r(r-r_1)}+\sqrt{(r-r_4)(r-r_3)}}{\sqrt{r(r-r_1)}-\sqrt{(r-r_4)(r-r_3)}} \right) \label{solTdiv}
\end{align}
\end{widetext}
where the last two terms diverge for $x=x_{\infty}$. We find the Taylor expansions of these terms as
\begin{align}
\frac{b\sqrt{R(r)}}{r-r_3} & = r+r_3+\mathcal{O}\left(\frac{1}{r}\right)\,, \label{DivTaylorlin}\\
2\ln(Z) & = 2\ln\left(\frac{2}{r_4+r_3}\right) + 2\ln r + \mathcal{O}\left(\frac{1}{r}\right)\,.\label{DivTaylorlog}
\end{align}

\bibliographystyle{unsrt}
\bibliography{biblio}

\end{document}